\documentclass[%
 reprint,
 amsmath,amssymb,
 aps,
]{revtex4-2}

\usepackage{graphicx}% Include figure files
\usepackage{dcolumn}% Align table columns on decimal point
\usepackage{bm}% bold math
\begin{document}

\preprint{APS/123-QED}

\title{Replacing the Notion of Spacetime Distance by the Notion of Correlation}% Force line breaks with \\

\author{Achim Kempf}
 \altaffiliation{Department of Applied Mathematics\\ University of Waterloo\\ 200 University Avenue West\\ Waterloo, Ontario, N2L 3G1, Canada}%Lines break automatically or can be forced with \\


\begin{abstract}
Spacetime is conventionally viewed as a stage on which actors, in the form of fields, move. Here, we explore what may lie beyond this picture. The starting point is the observation that quantum fluctuations of fields are the more strongly correlated the shorter their spacetime distance. As a consequence, the notion of spacetime distance, including the metric as its infinitesimal version,  can be replaced by the notion of correlation strength.  
This suggests a new picture in which abstract (2-point and multi-point) correlators are the primary structure. In general, these abstract correlators can only be described as information theoretic structures and, in principle, they need not bear any relationship to quantum fields or spacetimes. These correlators may allow approximations, however, so that, in certain regimes, they can be mathematically approximately represented as the 2-point and multi-point functions of quantum fields living on a spacetime. In this way, the standard picture of a curved spacetime with fields whose correlators arise from Feynman rules would merely be a convenient approximate picture while the underlying picture is that of a spacetime-less and field-less information-theoretic structure of abstract correlations. 

\end{abstract}

\maketitle

%\tableofcontents

\section{Introduction}

The discoveries of general relativity and quantum  theory each required the abandoning of major misconceptions. Today, the fact that it has turned out to be extremely hard to unify quantum theory and general relativity suggests that at least one more major misconception will need to be overcome. But what deeply-held belief about the nature of spacetime and matter may need to be abandoned to clear the path for the development of the theory of quantum gravity? 

One belief regarding spacetime and matter is that, while they do interact, they are fundamentally different, with spacetime representing a stage, which is itself dynamical, on which actors, in the form of massive or massless matter, move.
The present paper, which lays out ideas first presented orally in \cite{AK-vienna}, asks if the very stage-and-actors picture could be a misconception that needs to be abandoned, and it explores one possibility for what new picture lies beyond. 

\section{Probing general relativity's description of spacetime}

For inspiration, let us look for hints in some of the currently most successful descriptions of spacetime. 
One general relativistic description of a spacetime is as a pair, $(M,g)$, where $M$ is a differentiable manifold and $g$ is a Lorentzian metric. Equivalently, a spacetime is often described as a manifold with Christoffel symbols, $\Gamma$, or a connection $1$-form, $\omega$. Also equivalently, general relativity describes a spacetime as a pair $(M,\sigma)$, where $M$ is the differentiable manifold and $\sigma(x,x')=\frac{1}{2}\tilde{g}(x,x')$ is the Synge world function. Here, $\tilde{g}(x,x')$ is the geodesic distance between the events $x$ and $x'$, as far as that distance is unique. The Synge function, $\sigma$, contains all information about a spacetime since it allows one to recover its metric, \cite{synge,fulling,birrelldavies}: 
\begin{equation}
g_{\mu\nu}(x)=\lim_{x\rightarrow x'} \label{geo} \sigma(x,x')_{;\mu\nu} = %-\lim_{x\rightarrow x'} \sigma(x,x')_{,\mu,\nu'}=
-\lim_{x\rightarrow x'} \frac{\partial}{\partial x^\mu}\frac{\partial}{\partial x'{}^\nu} \sigma(x,x')
\end{equation}
For later reference, we notice in Eq.\ref{geo} that it suffices to know the bi-scalar function $\tilde{g}$ in an infinitesimal neighborhood of its diagonal. 
Finally, we should add that, to complete the general relativistic descriptions of spacetimes, Einstein also provided an exact mapping between the mathematical concepts and concrete physical measurements, based on rods and clocks \cite{einstein}. 

To search for hints at what may lie beyond general relativity, let us now probe general relativity for curiosities or odd features in the way that it describes spacetime. 
One curious feature of general relativity's description of spacetime is that it utilizes two different notions of distance. One notion of distance is that of coordinate distance. Finite coordinate distances are not covariant but infinitesimal coordinate distances are used, covariantly, to define the topology (in the sense of open neighborhoods) of the manifold, which in turn is used to define the notions of continuity and differentiability of the manifold and also to define the limit taking for derivatives and integrals. The second notion of distance is the geodesic distance. 

Further, this leads to the curious feature of general relativity that a spacetime manifold's topology (in the sense of open sets) is ignorant of the drama of the light cone. Points that are arbitrarily close neighbors with respect to the manifold's topology (in the sense of open sets) can be on either side of a light cone - which is the difference between being fully causally connected or not at all. 

Underlying the presence of two notions of distance in general relativity is another curious feature of general relativity. On one hand, general relativity describes space and time identically except for a minus sign in the metric. On the other hand, space and time are to be measured using rods and clocks - but rod-like and clock-like physical instruments appear to differ by substantially more than a minus sign. 
%Let us explore the role of rods and clocks further. 
It also appears odd that rods and clocks should play a fundamental role in general relativity, given that they each measure finite and therefore non-covariant coordinate distances. Further, Nature does not provide rods or clocks in Einstein's sense at sub-atomic scales. 

For now, let us take away the hint that it is worth considering to replace rods and clocks in general relativity in some way with tools that are more canonical. 

One may ask, for example, whether it is useful to replace traditional clocks that define a notion of time by counting periodic processes with clocks that measure a notion of time that is based on the exponential decay of unstable particles. Intuitively, this would amount to switching from clocks describable by the oscillations of a complex exponential function to a decay-based clock that measures time through a decay-describing real exponential  function, thereby possibly accounting for the extra sign in the metric's signature. Still, even such decay-based rather than oscillation-based clocks would appear to differ from rods in their physical appearance by more than a minus sign. We will here, therefore, not follow these lines. 

Instead, let us explore how both clocks, as well as rods, could be replaced by tools that are more canonical in the sense that they allow us to determine spacetime distances directly instead of inferring them from measurements of spatial and temporal distances which are themselves not covariant. This could provide us not only with a more canonical method for measuring the spacetime distance between two events. We should thereby also obtain a new method to map a spacetime's curvature. This is because as,  Eq.\ref{geo} showed, knowledge of the (even just infinitesimal) spacetime distances suffices to calculate the metric.     

\section{Measuring spacetime distances by means of  correlators}

There may exist multiple ways to replace rod-like and clock-like tools by more canonical tools for measuring spacetime distances. In the present paper, the idea that we pursue  is to replace rods and clocks by quantum field vacuum fluctuations. This is possible because the quantum fluctuations of a field are correlated and the strength of the correlation decays with the magnitude of the spacetime distance. The strength of the correlation can, therefore, serve as a measure of the spacetime distance.   

For an example of a correlator of the fluctuations of a quantum field, let us recall the Feynman propagator of a free massless scalar. In flat spacetime it reads:
\begin{eqnarray}
    G_F(x,x') & = & \langle 0\vert T\hat{\phi}(x)\hat{\phi}(x')\vert 0\rangle\\
    & = & -\int \frac{d^4p}{(2\pi)^4}\frac{e^{-ip_\mu (x^\mu-x'^\mu)}}{p_\mu p^\mu +i\epsilon}  \\
        & = & \frac{1}{4i\pi^2} \frac{1}{(x_\mu -x'_\mu)(x^\mu -x'^\mu)-i\epsilon}. 
\end{eqnarray}
Here, as well as in curved spacetime \cite{fulling,birrelldavies,brewster}, the correlator $G_F(x,x')$ is finite both inside and outside the lightcone. On the lightcone, it diverges and changes sign. 
Important for us here is that as we move away from the lightcone, i.e., as we increase the magnitude of the spacetime distance, the smaller the absolute value of $G_F(x,x')$, i.e., the weaker do the correlations become. Let us review  the reasons why the correlator decays away from the lightcone into both the timelike and the spacelike regions. 

First, inside the lightcone, the correlations are caused by the propagation of perturbations. In the course of the propagation, the perturbation spreads out and therefore generally weakens, at least initially. (For finite propagation distances in curved spacetime, there can be, for example, lensing-type and echo effects, see, e.g., \cite{ak-etal-bh}.) Hence, inside the lightcone, the correlator at least initially decays for increasing timelike distances. 

Second, the correlations outside the lightcone exist because the vacuum is a spatially entangled state. To see why the correlations decay away from the lightcone, i.e., here for increasing spacelike distances, let us recall the simple case of a quantized, massive Klein Gordon field. It can be viewed as consisting of one degree of freedom $\hat{\phi}(x,t)$ at each position, $x$. Holding a position $x$ fixed, its degree of freedom, $\hat{\phi}(x,t)$, would obey an independent quantum harmonic oscillator equation  
\begin{equation}
    \ddot{\hat{\phi}}(x,t) = - m^2 \hat{\phi}(x,t) 
\end{equation}
if it were not for the existence of the Laplacian term in the Klein Gordon equation: 
\begin{equation}
    \ddot{\hat{\phi}}(x,t) - \Delta \hat{\phi}(x,t)= - m^2 \hat{\phi}(x,t) 
\end{equation}
The Laplacian term couples spatially neighboring harmonic oscillators. Therefore, the ground state of these coupled harmonic field oscillators is an entangled state, hence the correlations.
Since the Laplacian couples only neighboring field oscillators, these correlations decay with the spacelike distance at a rate that is dimension dependent. In 1+3 dimensions, the decay is polynomial for massless fields and exponential for massive fields. This is the case at least for small distances. At large distances, on curved spacetimes, there may again occur, for example, analogs of lensing-type effects. See, e.g., \cite{valentini,reznik1,reznik2,sorkin,srednicki} for early work, \cite{jasonmaria} for more recent work and \cite{holoreview} for a recent review of the related topic of holography and quantum information.

So far, we established that a correlator of quantum field fluctuations, such as the Feynman propagator, $G_F(x,x')$, can be used as a measure of spacetime distances, or at least of small spacetime distances. From Eq.\ref{geo}, we know that knowledge of small spacetime distances, in the form of knowledge of the Synge function $\sigma(x,x')$ near its diagonal, is sufficient to calculate the metric. It was shown in \cite{per} that, similarly, knowledge of $G_F(x,x')$ near its diagonal suffices to calculate the metric tensor:
\begin{eqnarray}
g_{\mu\nu}(x) & = & -\frac{1}{2} \left(\frac{\Gamma(D/2-1)}{4\pi^{D/2}}\right)^\frac{2}{D-2} \nonumber \\ & & \times \lim_{x\rightarrow y} \frac{\partial}{\partial x^{\mu}}\frac{\partial}{\partial y^{\nu}} \left( G_F(x,y)^{\frac{2}{2-D}}\right)
\label{per}
\end{eqnarray}
Here, $D$ is the spacetime dimension. (The case $D=2$ is special and has a different expression, see \cite{per}.) 
The fact that the metric tensor can be calculated from the Feynman propagator can be seen also by this consideration:
Knowledge of the Feynman propagator implies knowledge of the lightcones because the propagator diverges and changes sign on the lightcones. But 
knowledge of the lightcones of a spacetime manifold determines the spacetime's metric tensor up to a conformal factor, see \cite{hawkingellis}. The propagator also provides the remaining conformal factor through its finite decay near the lightcone. 

For completeness, it is worth mentioning that the reconstruction of the metric from the Feynman propagator does not depend on the the vacuum state. This is important because different observers may identify different states as their vacuum state. 
Recall that in the absence of a unique vacuum state, Feynman propagators can differ by homogeneous solutions to their equation of motion, such as the Klein Gordon equation $\Box G_F=\delta/\sqrt(g)$. The metric is calculated, see Eq.\ref{per}, by differentiating a negative power of the Feynman propagator, i.e., by differentiating a positive power of the wave operator, $\Box$. The matrix elements of $\Box$ are independent of which homogeneous solution one may choose to define a Feynman propagator, i.e., a right inverse, $G_F$ of $\Box$. Concretely, in Eq.\ref{per}, any choice of $i\epsilon$ prescription for the propagator drops out because $i\epsilon$ prescriptions are in the denominator but since the propagator appears to a negative power, $\epsilon$ is in the numerator. Hence, the limit $\epsilon \rightarrow 0$ can be taken before using Eq.\ref{per} to calculate the metric. 

Our conclusion so far is that a classical spacetime, i.e., a Lorentzian manifold, can be viewed as a pair $(M,G_F)$ where $G_F$ is a Feynman propagator of a scalar field. 
A key difference between describing a spacetime using a pair $(M,\sigma)$ or a pair $(M,G_F)$ is that, in the former case, the traditional measurement of a geodesic distance requires the use of rods and clocks along the geodesic. In contrast, in the latter case, we replace rods and clocks, which are non-canonical human artifacts, by the naturally occurring fluctuations of a quantum field. The correlations in the quantum fluctuations of a field are sufficiently modulated by the underlying spacetime's curvature to allow us to reconstruct the metric of the spacetime. Similarly, it should be possible to use spinorial and tensorial Feynman propagators, after scalar contractions, to determine the metric.

In practice, the measurement of a field correlator, such as a Feynman propagator, would require, in principle, the detection and counting of quantum field fluctuations, a difficult notion. With present technology, quantum fluctuations of the vacuum of the electromagnetic field can be measured with some accuracy in table-top quantum homodyne detectors, see, e.g., \cite{ralph,milburn}. Quantum optical measurements of the correlations of spacelike or timelike separated electromagnetic quantum vacuum fluctuations may  become feasible in table-top experiments. In principle, with sufficient accuracy, such types of experiments could pick up gravity-caused modulations of the functional form of a Feynman propagator. From the Feynman propagator, the metric could then be calculated. In principle, therefore, such experiments, if mobile and sufficiently accurate, could be used to map the curvature of spacetime. In a more indirect sense, particle accelerators such as the LHC can be interpreted as devices that test Feynman rules and to determine the functional form of Feynman rules, including the Feynman propagators. 
With sufficient accuracy, a mobile particle accelerator could, therefore, also be used to map spacetime's curvature by determining a gravity-modulated Feynman propagator. 

To conclude, we arrived at the finding that instead of using rods and clocks to map a spacetime manifold's curvature, as Einstein envisaged,  we could map a spacetime manifold's curvature by measuring $G_F(x,x')$ close to its diagonal, i.e., by measuring the local correlations of quantum field fluctuations. This is because the Feynman propagator $G_F(x,x')$ can then be used to recover the traditional metric-based description of a spacetime through Eq.\ref{per}. Intuitively, this is possible because the strength of the correlations of the quantum field fluctuations that is encoded in the propagator is a proxy for the covariant distance, the correlations being the stronger the smaller the covariant distance.

\section{Replacing the notion of distance by the notion of correlation}

So far, we replaced one set of tools to map a spacetime's curvature with another set of tools to map a spacetime's curvature, namely by replacing rods and clocks with the correlator of quantum field fluctuations. We thereby assumed that there exists an underlying Lorentzian spacetime to be mapped.   
We now return to the main objective, which is to challenge the validity of the picture of a spacetime-stage that hosts matter-actors.

To this end, we begin by asking what if the reasonable assumption is true that, in Nature, there is no spacetime in the exact sense of a Lorentzian manifold? In this case, what we described above as the reconstruction of a spacetime from a Feynman propagator can only be approximate. 

This suggests to explore the possibility that the primary structure of Nature is not that of a Lorentzian spacetime-stage with matter-actors in the form of quantum fields but that instead the primary structure of Nature consists of abstract correlators. While the abstract correlators would describe all regimes of Nature, only in some regime, which may be called the `low energy' regime, the abstract correlators could be approximately represented in the sense that they could be viewed, at least approximately, as arising from quantum field fluctuations on a curved spacetime, as, e.g., in \cite{ak-jmp}. 

If abstract $2$-point and multi-point correlation functions are the primary structure, Nature would be information-theoretic in nature. 
The notions of a spacetime and matter would then be secondary in the sense that these notions only emerge in the low energy regime, as convenient notions for describing the structure of the approximate mathematical representations of the abstract correlators in terms of QFTs on a spacetime.

In the low energy regime, it would be the Feynman rules (to tree level) and ultimately the full sums of Feynman graphs of the standard model of particle physics on curved spacetime which would serve as a good approximate mathematical representation of the abstract correlators. At high energies, the abstract correlators would not possess a representation that makes them appear to arise from the quantum fluctuations of fields on a background Lorentzian spacetime. Not only would there be no notion of spacetime but also no notion of matter belonging to definite species of fields that could live on a spacetime. Instead, the abstract correlators would need to be thought of as mere structures that may be best described information theoretically.  

Here, we can not answer the question what determines the structure of the abstract correlators, 
as this question is as hard as asking what determines the dimension of spacetime, the structure of the standard model field content and interactions, and what lies beyond.

Let us now return to our aim to challenge the widely-held  picture of a spacetime-stage and matter-actors, in order to perhaps get a glimpse of a possible new picture that could lie beyond. 
To this end, let us consider how, in this new picture, the derived notions of a spacetime stage and matter actors would be seen as breaking down towards the Planck scale, from the perspective of an experimenter who approaches the Planck scale from low energies: correlators, such as a propagators, should become less and less knowable at high energies and small distances. For example, to measure a correlator, such as a Feynman propagator, with some accuracy requires, in principle, a large number of measurements since statistics needs to be accumulated to obtain a reliable value for a correlator. Repeated measurements can be spaced out in small regions of spacetime but as these regions are chosen smaller (speaking in the conventional picture), interactions increase, significant renormalization is needed and eventually a natural ultraviolet cutoff may arise - limiting the knowability of the statistics of the quantum fluctuations. 
If so, from the perspective of the traditional picture of a spacetime-stage with matter-actors, the Planck scale would not be a regime of exotic phenomena or of wild quantum fluctuations of spacetime and matter. Instead, the Planck scale might appear as a regime of poor statistics. The statistics of the correlators, or Feynman propagators, would be too poor to even approximately assign a classical metric. (Crudely, the notion of a classical spacetime arising approximately, in some regime, from abstract correlators could be compared to the notion of a classical path arising approximately, in some regime, from a quantum particle's wave function.) From the information-theoretic perspective of the abstract correlators, this phenomenon of inaccessibility of information in the ultraviolet may appear, for example, as a bandlimitation for matter fields which in turn induces a corresponding `bandlimitation' on the knowability, by means of matter-based measurements, of spacetime curvature. See, e.g., \cite{shannon,ak-shannon1,ak-shannon2,ak-shannon3,ak-shannon4,ak-shannon5,ak-shannon6,ak-shannon7,ak-shannon8}.

\section{Identifying the geometric degrees of freedom} 

We have arrived at a picture in which correlators, such as a Feynman propagator, are primary, with a metric spacetime and quantum fields emerging as derived, approximate concepts that provide a useful language for the `low energy' regime. 
Let us for now focus on this low energy regime, defined as the regime where both the metric-based and the correlator-based descriptions are valid. 

In this low energy regime, we can now identify a problem that persists when transitioning from the metric-based picture to the new abstract correlator-based picture: the problem is that a correlator, such as a Feynman propagator, $G_F(x,y)$, is still a function of arbitrary parameters, much like the metric. The propagator is, therefore, encoding its geometric information highly redundantly. This is because, much like the metric, the affine connection, and the Synge function, the functional form of a propagator changes under diffeomorphisms. This leaves us with the task to mod out the diffeomorphism group if we wish to isolate the geometric degrees of freedom, i.e., to identify the Lorentzian structure. 

For a first attempt at extracting the diffeomorphism invariant information contained in a Feynman propagator, let us start by recalling that, functional analytically,  the Feynman propagator is a right inverse \begin{equation}
    WG_F=\delta
\end{equation} of its wave operator, $W$, such as  
\begin{equation}
    W:=\sqrt{g}(\Box +m^2).\label{wave}
\end{equation}
    The wave operator is a self-adjoint operator which, as we discussed above with Eq.\ref{per}, inherits all geometric information from the propagator. As a self-adjoint operator, the wave operator possesses a real spectrum, $spec(W)$, and this spectrum at first sight appears to be what we are looking for, namely a set of invariants under the diffeomorphism group. We arrive at Lorentzian spectral geometry, see, e.g., \cite{yasamanmarco,kac,weyl,datchev}, the discipline that asks: to what extent does the spectrum of a wave operator determine a Lorentzian manifold? 

In fact, $spec(W)$ is not a large enough set of invariants to identify the Lorentzian manifold. There are two basic reasons. The first reason is that the spectra of the typically hyperbolic wave operators tend to be continuous and therefore particularly information poor. The spectra of hyperbolic wave operators can be made discrete with suitable infrared cutoffs. This necessitates choices of boundaries and boundary conditions. These, however, are strongly affecting the resulting spectra and thereby obscuring the extraction of geometric information from the spectra. 
The second reason for why $spec(W)$ is not a large enough set of invariants to identify a Lorentzian manifold, even if suitably discretized via an IR cutoff, is more fundamental. The reason is that $spec(W)$ is a set of invariants under the action of the entire unitary group in the function space, which is a larger group than the diffeomorphism group since it also contains, for example, Fourier transforms. This means that $spec(W)$ can be and generally is smaller than the set of invariants under only the diffeomorphism group. 

As an aside, let us briefly discuss an approach \cite{ak-shannon4,bhamre} to overcoming this problem  by introducing the tool of  infinitesimal spectral geometry. Conventional spectral geometry aims to solve the highly nonlinear problem of determining to what extent the spectrum of an operator on a manifold determines the manifold's metric. Infinitesimal spectral geometry aims to solve the simpler linear problem of determining to what extent an infinitesimal change of the spectrum of an operator on a manifold determines the corresponding infinitesimal change in the manifold's metric. Infinitesimal changes are then iterated to obtain finite changes of the curvature of the manifold from finite changes of the spectrum, as far as well defined. This approach yields a new perspective for why the set of geometric invariants $spec(W)$ is generally incomplete: Only in two dimensions is the metric essentially scalar. In higher dimensions, perturbations of the metric are truly tensorial and therefore cannot be covariantly expanded in the eigenbasis of a scalar wave operator. This suggests, as a remedy, to work with the spectra not of scalar wave operators but of wave operators of covariant symmetric 2-tensors since this will guarantee that any small change in the spacetime metric can be covariantly expanded in the eigenbasis of the wave operator. To this end, Feynman propagators of spin-2 particles that are composites  could be used, as gravitons are the only expected non-composite spin-2 particles.

We now propose a new approach to extracting the geometric, diffeomeorphism invariant information from the abstract correlators. 
To this end, let us retrace our steps below Eq.\ref{wave}. 
We started with the knowledge that the Feynman propagator $G_F$ as well as its wave operator $W$  contain the complete information about the metric if given in the position representation. Due to diffeomorphism invariance, they do so in a highly redundant way. We therefore considered the spectrum of the wave operator, since it consists of diffeomorphism invariants that carry geometric information - though not the complete set of geometric information. In other words, we observed that while  the wave operator contains all geometric information when given in a position basis, it does not contain the complete geometric information when given in its eigenbasis, i.e., when we only know its spectrum. 

This tells us that knowing the wave operator in its eigenbasis (i.e., knowing nothing but its spectrum) and in addition also knowing a unitary transformation from that eigenbasis to a position basis is sufficient to calculate the metric. This is because we can then transform the Feynman propagator or wave operator from the eigenbasis of the wave operator into a position basis and from there arrive at the metric using Eq.\ref{per}. 

The question is, therefore, if we can find such a unitary transformation on the basis of knowing only the abstract correlators. The answer is yes. To see this, recall that, so far, we have only utilized the information contained in the abstract 2-point correlators, i.e., in the Feynman propagators. The abstract $n$-point correlators for $n>2$, which we have not yet used, happen to contain exactly the information that is needed to calculate unitaries that map from the eigenbasis of a propagator's wave operator to a position bases. The reason is that these multi-point correlators describe the vertices of interactions. And these vertices are local. 
This means that from the abstract correlators in an arbitrary basis, we can always calculate unitary transformations to position bases, namely by diagonalizing these vertices (as operators from a $n$-fold tensor product of the space of fields into itself, with $n$ depending on the valence of the vertex). Only in the position basis are the vertices of the Feynman rules of a local quantum field theory diagonal, i.e., only in a position representation are the vertices proportional to products of Dirac deltas. For example, the $3$-vertex of $\lambda \phi^4$ theory is usually given in the momentum basis but it can be expressed in any basis of the space of fields, for example, in a Bargmann Fock basis, see \cite{ak-ucr-qft}. In a position basis, and only in a position basis, is the vertex diagonal in the sense that it takes the form \begin{equation}
    V(w,x,y,z)=-i\lambda \delta(w-x)\delta(y-z)\delta(w-z)
    \end{equation}
which expresses the locality of the interaction.

In conclusion, we have, therefore, arrived at a new method obtain the metric from  correlation functions, namely here from knowledge of a propagator and a vertex of a QFT. Crucially, the new method can be used if the propagator and vertex are given in any arbitrary basis in the function space or also if they are given basis independently. Given the propagator and vertex, the method consists in determining a basis in which the vertex is diagonal (a position basis, therefore), then transforming the propagator into that basis and finally deriving the metric from the propagator. Unlike infinitesimal spectral geometry, the new method, therefore, works straightforwardly for spacetimes of any dimension and signature. As for the assumption of the diagonalizability of the vertex, as far as is known, the vertices of all physical theories are local and therefore diagonalizable (i.e., possess representations as products of Dirac deltas), at least in the low energy regime, which is where we are here deriving a metric from the correlators. The dissolution of the notion of a spacetime manifold as one approaches the Planck scale can then manifest itself mathematically as the non-diagonalizability of the vertex, i.e., in the appearance of nondiagonal terms in the vertex correlators in any basis.

We remark that we can now see from a new perspective how, for example at the Planck scale, abstract $n$-point correlators can fail to possess a representation in terms of a quantum field theory whose interactions are local and which lives on a classical curved spacetime. This happens in regimes where the $n$-point correlators no longer admit even an approximate diagonalization. 

The new approach to extracting the geometric information from $n$-point correlators can be viewed as a generalization of  spectral geometry: conventional spectral geometry studies to what extent the shape of a manifold can be inferred from the spectrum of a wave operator of a free field that lives on the manifold. The new approach is to consider not free but interacting  fields on the manifold, or even just one field that is self-interacting, e.g., though a $\hat{\phi}^4$ interaction. This yields nontrivial $n$-point correlators for $n\ge 2$. Independently of the basis in which these Feynman rules are given, they contain the basis-independent information of a) the spectrum of the propagator and b) the changes of basis from the eigenbasis of the propagators to the position bases, defined as those bases in which the   vertices are diagonal, i.e., local. Together, these two sets of basis independent and therefore also diffeomorphism independent information form a complete set of invariants to describe a metric manifold. 

It should be interesting to see if, in acoustic spectral geometry, this translates into the ability to hear the shape of a thin curved vibrating object if drumming it weakly as well as strongly enough to invoke nonlinear oscillations.  

\section{Outlook} 
There are, of course, open questions regarding the picture in which abstract correlators are primary, with the conventional picture of a spacetime stage that hosts matter actors only emerging in certain regimes as useful but approximate representations of the abstract correlators. 
For example, one may ask whether 
there are new prospects for deriving the dimensionality of spacetime and, regarding dimensionality, what the relationship to holography could be. 

It is worth considering here the fact that  any first quantized or suitably UV and IR regularized second quantized theory formulated in one number of spatial dimensions can be unitarily mapped into an equivalent first or second quantized theory in any other chosen number of spatial dimensions. The reason is that the Hilbert spaces of first quantized theories with a finite number of degrees of freedom are separable, i.e., they possess countable Hilbert bases. Quantum field theories, after suitable ultraviolet and infrared cutoffs, also possess only a finite number of degrees of freedom and their Hilbert spaces are, therefore, also separable. All separable infinite-dimensional Hilbert spaces, however, are unitarily equivalent. (We are assuming here that the regularizations are not so drastic that they reduce the dimension of the Hilbert spaces to finite numbers since finite-dimensional Hilbert spaces are unitarily equivalent only if their finite dimensions match.) 

For example, using Cantor's diagonal counting, the countable eigenbasis of a 1-dimensional harmonic oscillator can be unitarily mapped into the also countable Hilbert basis of a 2-dimensional harmonic oscillator, or, e.g., into the  countable eigenbasis of a hydrogen atom in a three dimensional box. Of course, what is local in one theory will generally not be local in the unitarily equivalent theory. 
Similarly, the equivalence of a regularized second quantized theory in one number of spatial dimensions to a regularized second quantized (or first quantized!) theory in an arbitrary different number of spatial dimensions, is guaranteed, i.e., it is not special per se. What can make such an equivalence special is if the two equivalent theories in question are each of interest in their own right. 

From the perspective of the picture where abstract correlators are primary, the determination in which regime the correlators can be represented as arising, approximately, as the correlations of quantum fluctuations of fields on a spacetime, tells us in effect what dimension of spacetime and what matter content the given abstract correlators describe in some regime. 
In order to investigate these questions, a technical challenge will be to develop functional analytic methods to describe Feynman rules, for example, those of a scalar $\hat{\phi}^4$ theory, basis independently, for example, in terms of the spectra of wave operators and the unitaries that map the eigenbasis of a wave operators into bases in which the vertices are (essentially) diagonal. This analysis should help identify those functional analytic properties of the vertices, i.e., of the $n$-point correlators, that determine the regime in which they can be viewed as at least approximately diagonalizable. Knowing those functional analytic properties could help explore possible structures that determine the abstract correlators. 

Presumably, the natural language to study questions about the abstract correlators is information theory. For example, as we briefly discussed, the presence of a natural ultraviolet cutoff could manifest itself in the abstract correlators as a form of bandlimitation, in which case generalized Shannon sampling theory,\cite{ak-shannon1,ak-shannon2,ak-shannon3,ak-shannon4,ak-shannon5,ak-shannon6,ak-shannon7,ak-shannon8,ak-soda2}, which is related to minimum length uncertainty principles \cite{ak-ucr1,ak-ucr2,fabio}, could provide useful tools.

Among the many open questions is also how to interpret   what 
in the conventional picture
are widely-separated entangled systems. In the new picture, where spacetime distance is, by definition, inferred from correlations, such systems would appear to be `close' by definition, i.e., they may be considered `close' without needing an appeal, for example, to conventional wormholes, see also \cite{worm}.   

From the new perspective where abstract correlators are primary, it should also be interesting to explore possible links to candidate quantum gravity theories, see, e.g., \cite{loops,loll} and also \cite{ding} which is based on the Synge function  and quantum indefinite causal structures, see, e.g., \cite{ics}, as well as to studies that aim to link the structure of the standard model to algebraic structures and discrete spacetime models, see e.g., \cite{cohl,connesSM}. Of relevance here could also be the in-depth investigations into the relationship between the possible dynamics of matter and the correspondingly possible dynamics of gravity, in \cite{schuller}. 
It will be interesting as well to explore possible connections to the physics and formalisms of quantum reference frames and related studies of notions of distance via correlations between observables, see, e.g., \cite{qrf,gia2,est}.

Finally, our perspective where abstract correlators are primary is philosophically close to various approaches, such as relational quantum mechanics, \cite{rovelli-relational} and, in particular, 
\cite{mermin}. The approach in \cite{mermin} is not concerned with quantum field theory or curved spacetime. However, its central observation could be of interest also to our approach here: In \cite{mermin}, Mermin showed that the set of correlators among any chosen complete set of subsystems of a system provides a complete tomography of the system's state - hence the motivation there to consider correlators as primary. It should be interesting to explore how or to what extent this result can be applied to quantum field theories, although limitations to localizability, as e.g., described by Malament's theorem, make the consideration of localized subsystems difficult in quantum field theory, even in the low energy regime. Worth mentioning here are also attempts at describing Nature information theoretically based on the idea of zeroth, first and second quantizing the notion of a binary alternative, see, e.g., \cite{vWeiz,revvW}.

\section*{Acknowledgments}
The author acknowledges 
support through a Discovery Grant of the National Science and Engineering Council of Canada (NSERC), a Discovery Project grant of the Australian Research Council (ARC) and a Google Faculty Research Award. The author is grateful for feedback from Barbara \v{S}oda, Jason Pye, Maria Papageorgiou,  Flaminia Giacomini and two anonymous referees. 
%The author is grateful for valuable feedback from Barbara \v{S}oda, Vivishek Sudhir, Jason Pye, Maria Papageorgiou and Flaminia Giacomini. 

\bibliographystyle{frontiersinSCNS_ENG_HUMS} % for Science, Engineering and Humanities and Social Sciences articles, for Humanities and Social Sciences articles please include page numbers in the in-text citations

\end{document}